\documentclass{article}
\usepackage{ijcai24}

\usepackage{times}
\usepackage{soul}
\usepackage{url}
\usepackage[hidelinks]{hyperref}
\usepackage[utf8]{inputenc}
\usepackage[small]{caption}
\usepackage{graphicx}
\usepackage{amsmath}
\usepackage{amsthm}
\usepackage{booktabs}
\usepackage[switch]{lineno}

\usepackage{listings}
\usepackage{enumerate}
\newtheorem{definition}{Definition}
\def\method{STET}
\usepackage{multirow}
\usepackage{multicol}
\usepackage{newfloat}
\usepackage{dsfont}
\usepackage{booktabs}
\usepackage{amssymb}
\usepackage[linesnumbered,ruled,vlined]{algorithm2e}

\title{Revisiting Noise Resilience Strategies in Gesture Recognition: Short-Term Enhancement in Surface Electromyographic Signal Analysis}

\begin{document}


\author{
Weiyu Guo$^1$
\and
Ziyue Qiao$^2$\and
Ying Sun$^1$\And
Hui Xiong$^1$\\
\affiliations
$^1$The Hong Kong University of Science and Technology, Guangzhou\\
$^2$Great Bay University, Dongguan\\
\emails
\{guoweiyu96, ziyuejoe\}@gmail.com,
\{yings, xionghui\}@ust.hk
}

\maketitle

\begin{abstract}

Gesture recognition based on surface electromyography (sEMG) has been gaining importance in many 3D Interactive Scenes. However, sEMG is easily influenced by various forms of noise in real-world environments, leading to challenges in providing long-term stable interactions through sEMG. Existing methods often struggle to enhance model noise resilience through various predefined data augmentation techniques.
In this work, we revisit the problem from a short term enhancement perspective to improve precision and robustness against various common noisy scenarios with learnable denoise using sEMG intrinsic pattern information and sliding-window attention. We propose a Short Term Enhancement Module(STEM) which can be easily integrated with various models. STEM offers several benefits: 1) Learnable denoise, enabling noise reduction without manual data augmentation; 2) Scalability, adaptable to various models; and 3) Cost-effectiveness, achieving short-term enhancement through minimal weight-sharing in an efficient attention mechanism.
In particular, we incorporate STEM into a transformer, creating the Short Term Enhanced Transformer (STET).
Compared with best-competing approaches, the impact of noise on STET is reduced by more than 20\%. We also report promising results on both classification and regression datasets and demonstrate that STEM generalizes across different gesture recognition tasks.

\end{abstract}

    \section{Introduction}

\begin{figure}[tbp] 
    \centering 
    \includegraphics[width=0.45\textwidth]{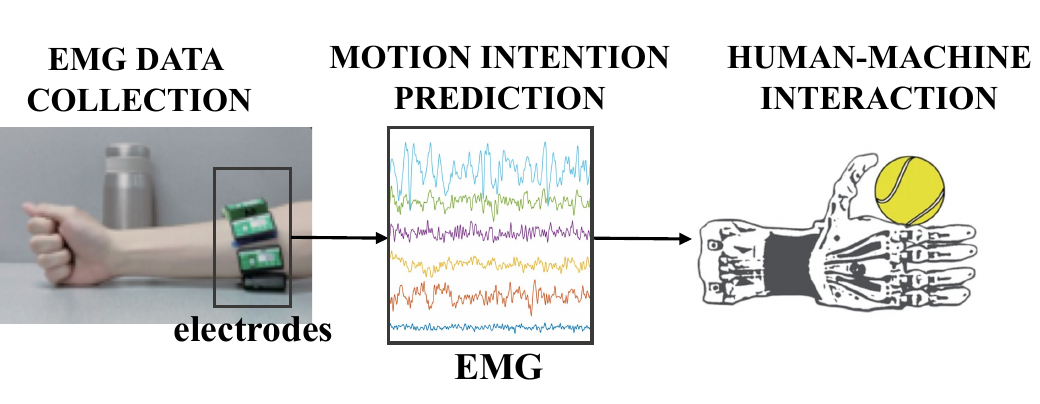} 
    \vspace{-4mm}
    \caption{Schematic Diagram of EMG-based Human-Computer Interaction System.} 
    \label{systerm} 
    \vspace{-4mm}
    \end{figure}

    Surface Electromyographic (sEMG) is a non-invasive technique for motoring muscle neurons firing, which is an effective way to capture human motion intention and has shown great application potential in the field of human-computer interaction (HCI) \cite{surveyEMGbasedHCI,liu2021neuropose,liu2020finger}. A schematic diagram of the EMG-based HCI System is shown in Figure \ref{systerm}. Compared to traditional HCI channels, sEMG has the advantages of being generated prior to actual motion (50-150 ms), containing rich motion intention information, and being easy to collect \cite{sun2020intelligent}. Therefore, there has been increasing interest in exploring EMG-based motion track \cite{liu2021wr} and pathological analysis.
    
    By treating sEMG as time series, deep sequential models \cite{Bi2019,tcn,becker2018touchsense,li2021gesture,du2017semi} have been applied to sEMG modeling. For example, Zhang \textit{et al.} employs a multi-task encoder-decoder framework  improve the robustness of sEMG-based Sign Language Translation (SLT) \cite{zhang2022wearsign}. Rahimian \textit{et al.} employs Vision Transformer (ViT)-based architecture (TEMGNet) to improve the accuracy of sEMG-based myocontrol of prosthetics \cite{rahimian2021temgnet}.  Although these methods demonstrate enhanced performance compared to traditional approaches, they process sEMG signals as generic time-series data, without specifically tailoring their design to address the unique characteristics of sEMG, such as their high variability and sensitive to external noise and interference. This oversight leads to issues such as difficulty in handling low signal-to-noise ratios due to changes in skin surface conditions and signal interference, and failing to capture subtle but important motion information in sEMG signals. As a result, the robustness and accuracy of existing models remain significantly challenged.
    
    Processing sEMG signals is challenging due to the complex noise mixed in the skin's surface and the presence of patterns across various time scales. Existing works, mainly focusing on long-term sequences, have used transformers to treat sEMG as a typical time series, aiming to enhance long-term dependencies. These approaches overlook the critical features which present in short-time scales. Short-time scale features are important in sEMG analysis, as they aid in distinguishing subtle movements and facilitate the removal of variable noise. For example, gestures like Index Finger Extension (IFE) and Middle Extension (ME), while similar in global sEMG patterns, can be differentiated through localized short-term signal variations.
    
   To this end, in this paper, we present a lightweight but powerful Module called Short-Term Enhanced Module (STEM) which utilizes sliding window attention with weight sharing to capture short-term features. Building on STEM, we further propose the Short-Term Enhanced Transformer (STET). STET leverages STEM to capture local signal changes, enhancing noise resistance, and then combines STEM with long-term features, further improving accuracy for downstream predictions. Furthermore, to enhance model robustness with minimal annotation, we propose a self-supervised paradigm based on sEMG Signal Masking to leverage the inherent variability in sEMG signals. 
    
Finally, we conducted extensive experiments on the largest public sEMG datasets. The experimental results show that STET surpasses existing methods by a significant margin in both gesture classification and joint angle regression tasks for single-finger, multi-finger, wrist, and rest gestures. Meanwhile, STET achieves strong robustness even when trained on pure data and tested on noisy data. Compared with best-competing approaches, the impact of noise on STET is reduced by more than $20\%$. Moreover, through visualizations, we show that the long-term and short-term features are complementary in sEMG-based gesture recognition tasks, and the fusion of the two features can make the classification boundary more obvious. This clearly demonstrates that short-term information is critical for sEMG-based gesture recognition and will provide a new design paradigm for future sEMG model design. In particular, we have deployed STET as an important functional component in our HCI system, which can offer a more intuitive and effective experience. Our real-world deployment is shown in the appendix.

    To the best of our knowledge, we are the first to highlight the short-term features in sEMG-based gesture recognition. Our contributions can be summarized as follows:
    \begin{enumerate}[1)]
    \item From the perspective of enhancing short-term features, we propose STEM, a learnable, scalable, and low-cost noise-resistant module. The integration of STEM into various neural networks has resulted in a marked improvement in their performance; 
    \item we introduce sEMG Signal Masking to self-supervised sEMG Intrinsic Pattern Capture Module to leverage the inherent variability in sEMG; 
    \item we conduct experiments on the largest wrist sEMG dataset, showing that our proposed method outperforms existing approaches in terms of accuracy and robustness. And short-term enhancement can be extended to other models like Informer.
    \end{enumerate}
    \section{RELATED WORK}
    \textbf{The EMG-based Intention Prediction of Human Motion} can be broadly divided into model-based and data-driven methods. Model-based methods typically combine disciplines such as kinesiology, biomechanics, and human dynamics to explicitly model the relationship between EMG and outputs (such as joint angles and forces). The model often includes specific parameters, such as joint positions and bone-on-bone friction, that need to be repeatedly experimented with and adjusted until the desired performance is achieved. In terms of parameter selection and determination, model-based methods can be further divided into kinematic models \cite{borbely2017real}, dynamic models \cite{koike1995estimation,Koirala2015,liupu}, and muscle-bone models \cite{wanglin2002,zhao2020emg,yao2018adaptive}.  Clancy \textit{et al.} used a nonlinear dynamics model to identify the relationship between constant posture electromyography and torque at the elbow joint \cite{Clancy2012}. Hashemi \textit{et al.} used the Parallel Cascade Identification method to establish a mapping between forearm muscles and wrist forces \cite{Hashemi2012}. However, model-based methods have a large number of parameters that are difficult to measure directly. Currently, only simple motion estimation with a limited number of joints and degrees of freedom is possible. In contrast to model-based approaches, data-driven methods do not require the measurement of various parameters. Recently, some researchers have begun to use temporal deep learning models to extract motion information from sEMG \cite{lin2022bert,zhang2022wearsign,Guo2021}. Lin \textit{et al.} proposed a method based on the BERT structure to predict hand movement from the Root Mean Square (RMS) feature of the sEMG signal \cite{lin2022bert}. Rahimian \textit{et al.} proposed a novel Vision Transformer (ViT)-based neural network architecture to classify and recognize upper-limb hand gestures from sEMG for use in myocontrol of prostheses \cite{rahimian2021temgnet}. However, these methods have neglected the modeling of short-term dependencies and have not considered the inherent variability in sEMG signals.

\vspace{-0.15cm}
\section{Preliminaries}

\subsection{Dataset}
\label{dataset}
\vspace{-0.15cm}
We conduct the experiments on Gesture Recognition and Biometrics ElectroMyogram (GRABMyo) Dataset \cite{pradhan2022multi}, which is the largest known open-source wrist EMG dataset with 43 subjects and has great potential for developing new generation human-machine interaction based on sEMG. 


\emph{Data processing.} The subjects performed 17 gestures of hand and wrist (including a rest period sEMG) according to the prompts on the computer screen. Each gesture was repeated 7 times, each lasting 5 seconds. 
In order to improve the convergence speed of the model, we use two methods (Max-Min normalization, $\mu$-law normalization) to normalize the data \cite{rahimian2020xceptiontime,recommendation1988pulse}. 
After normalization, we use a time-sliding window to split samples. In this paper, we set the window size as 200ms, and the overlap of adjacent windows is 10ms. $\mu$-law normalization can logarithmically amplify the outputs of sensors with small magnitudes, which results in better performance than linear normalization.

\begin{definition}[sEMG Signal Sequence]
An sEMG signal sequence is defined as a temporal signal sequence sampled by multiple sensors from a human wrist, which can be formulated as $\mathbf{X} = [\mathbf{x}_1, \mathbf{x}_2, .., \mathbf{x}_t]$, where $t$ is the time window. $\mathbf{x}_i = [x_{i,1}, x_{i,2}, ...x_{i,c}]$ represents the signal vector of $c$ sensors, where $x_{i,j}$ is the signal value of the $j$-th sensor in the $i$-th time step.
\end{definition}


\section{Technology Detail}

\subsection{Model Overview}
\begin{figure*}[htbp] 
\centering 
\includegraphics[width=0.85\textwidth]{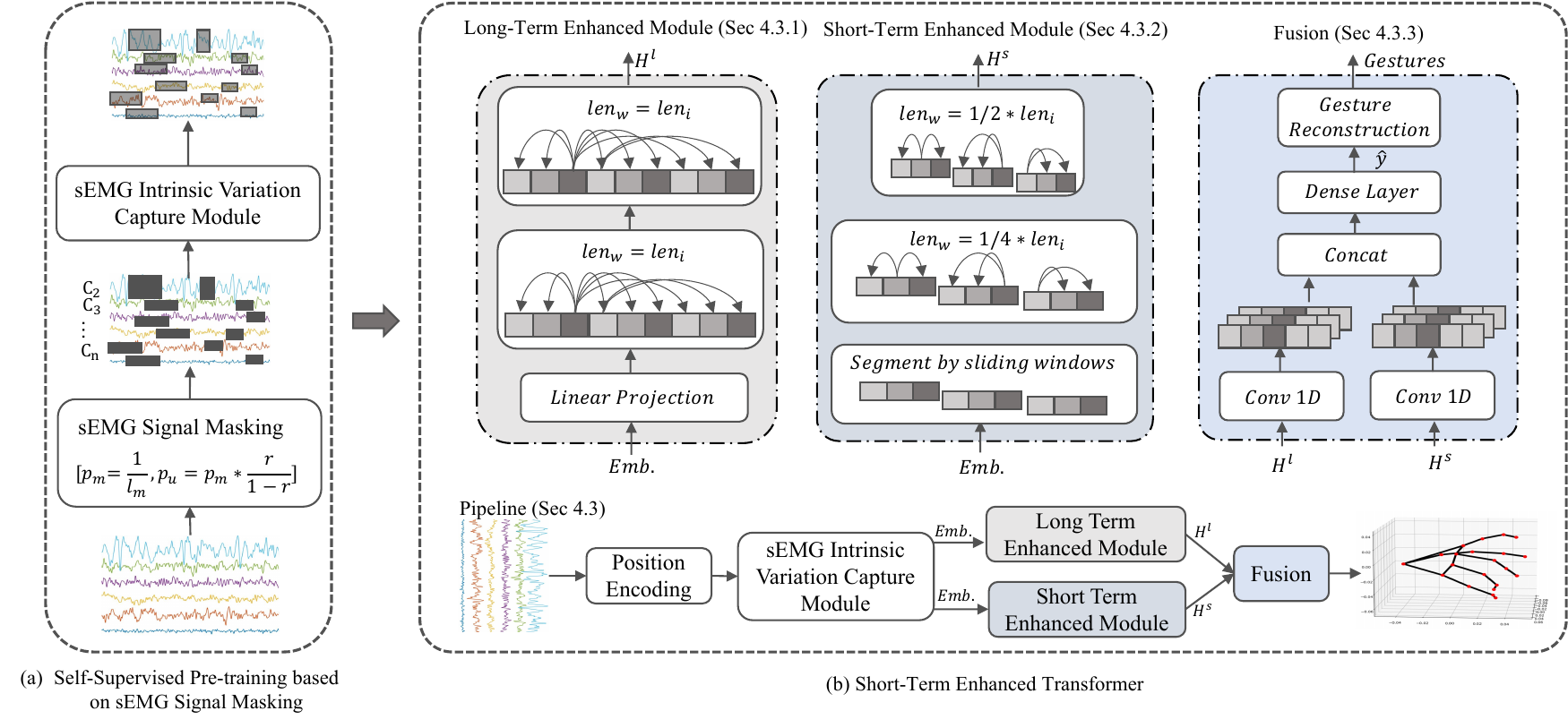} 
\vspace{-0.4cm}
\caption{Overview of STET. The sEMG signal is encoded using the sEMG Intrinsic Pattern Capture module, which is first pre-trained via sEMG signal Masking. A long-term and short-term enhanced module improves sEMG representations. An asymmetric optimization strategy addresses biases and imbalances in gesture recognition through an asymmetric classification loss.} 
\label{overview} 
\end{figure*}

Figure \ref{overview} illustrates the overview of our proposed framework for gesture recognition, which contains three components: (1) The \textit{sEMG Intrinsic Pattern Capture} module encodes the sEMG signal sequence into the hidden sEMG representations. A pre-training model with a segment masking strategy and MSE reconstructing loss is proposed to learn inherent variability from the sEMG signals into the model's parameters.
(2) The \textit{Long-term and Short-term Enhanced} module uses two decoupling heads to extract the long-term and short-term context information separately, which improves the sEMG representations in preserving both the global sEMG structure and multiple local signal changes of the sEMG.
(3) The \textit{Asymmetric Optimization} strategy addresses the problems of sample biases and imbalance in gesture recognition via an asymmetric classification loss, which can make the model focus on hard and positive samples to improve the recognition.

\subsection{sEMG Intrinsic Pattern Capture Module
}

\subsubsection{sEMG Signal Encoding}

Given the sEMG signal sequence $\mathbf{X} = [\mathbf{x}_1, \mathbf{x}_2, .., \mathbf{x}_t]$, we first project each signal $\mathbf{x}_i\in \mathbb{R}^c$ into a hidden embedding via a transformation matrix and add each signal embedding with an absolute position embedding. 
Then, we feed the output sequence into a $L$-layer Transformer and obtain the output signal embeddings $\mathbf{X}^{(L)} = [\mathbf{x}^{(L)}_1, \mathbf{x}^{(L)}_2, .., \mathbf{x}^{(L)}_t]$, which incorporate temporal context signal information for each position in the sequence.

\subsubsection{sEMG Signal Masking}
After the sEMG signal-extracting module is constructed,
we aim to use pre-training to exploit the intrinsic pattern and temporal semantics disclosed by the unlabeled sEMG signals (labeling sEMG is time-consuming and labor-intensive) and give a good initialization for the model parameters, then avoid the model focusing on some noisy features in the supervised learning task so as to over-fitting on some local minimums. Thus, we propose a sEMG Intrinsic Pattern Capture based on a signal masking strategy.

Specifically, given a transformed signal embedding sequence $\mathbf{X} = [\mathbf{x}_1, \mathbf{x}_2, .., \mathbf{x}_t]$, instead of adding masks on the sequence in terms of time steps like BERT, we add sensor-wise masks for the signal sequence of each sensor similar with~\cite{zerveas2021transformer}, which can encourage the model to learn more fine-grained temporal context dependency on the signal sequence of multiple electrodes. 
For the signal sequence of the $i$-th sensor, formulated as $[x_{1,i}, x_{2,i},...,x_{t,i}]$, i.e, the $i$-th column of $\mathbf{X}$, we generate a binary mask vector $\mathbf{m}_i\in \mathbb{R}^{t}$, where average $r$ radio of elements in $\mathbf{m}_i$ should be 0.15. 
Randomly generating $\mathbf{m}_i$ may cause a lot of isolated-masked signals, meaning one masked signal whose adjacent signals are unmasked. However, a single signal can be easily predicted by its immediately preceding or succeeding signals, making self-supervised learning easy to fitting on ineffective patterns and poor for learning temporal semantic information.
In consideration of this, we introduce a more complex masking strategy that aims to generate multiple masked segments on the sequence with an average length $l_m$, which means $m_i$ is composed of contiguous masked segments and unmasked segments. The length of masked segments follows a geometric distribution with mean $l_m$, and the length of unmasked segments follows a geometric distribution with mean $l_u$. Also, the $\frac{l_m}{l_u} = \frac{r}{1-r}$ so that the number of masked elements would follow the proportion $r$. 
The pseudocode of the masking algorithm is presented in Algorithm \ref{algorithm1}.

\begin{algorithm}
\caption{The Algorithm of sEMG Signal Masking}  
\label{algorithm1}
\LinesNumbered 
\KwIn{The length of the input signal sequence $t$,\newline
The number of signal sensors $c$,\newline
The average length of masked segments $l_m$,\newline
The masked radio $r$.
}
\KwOut{The mask matrix $\mathbf{M}$.}
\For{$i=1,...,c$}{
$\mathbf{m}_i = [True] * h$;\\
$p_m = \frac{1}{l_m}$; \tcp{probability of each masking segment stopping.} 
$p_u = p_m * r / (1 - r)$; \tcp{probability of each unmasked segment stopping.}
$\mathbf{p} = [p_m, p_u]$;\\
$state = Bool(random(0,1) > r)$; \tcp{the first state.}
\For{$j=1,...,t$}{
$\mathbf{m}_{i,j} = state$;\\
\If{$random(0,1) < \mathbf{p}[state]$}
{$state = \lnot state$;} 
}
}
\textbf{Return} $\mathbf{M} = \left([\mathbf{m}_i]_{i=0}^{c}\right)^T$;\\
\end{algorithm}


Then, we can mask the input sEMG signal sequence $\mathbf{X}$ by $\widehat{\mathbf{X}} = \mathbf{X} \odot \mathbf{M}$, where $\odot$ is elementwise multiplication and $\widehat{\mathbf{X}}$ is the masked input.
With the proposed Transformer-based sEMG signal encoder, we can obtain the output $\widehat{\mathbf{X}}^{(L)} = [\widehat{\mathbf{x}}_1^{(L)}, \widehat{\mathbf{x}}_2^{(L)}, ..., \widehat{\mathbf{x}}_t^{(L)}]$. 
For self-supervised learning, we add a linear layer on the top of masked output to reconstruct each sEMG signal $\widehat{\mathbf{x}}_i^{(L)}$ as $\widetilde{\mathbf{x}}_i \in \mathbb{R}^c$, 
which is the reconstructed sEMG signal in the $i$-th time step generated from the masked input.
Then, we minimize the Mean Squared Error (MSE) of the reconstructed signals and original signals on the masked positions for each sample:

\begin{equation}
\min \frac{1}{|\mathbf{M}|}\sum_{i=0}^{t}  \sum_{j=0}^{c} \mathds{1}(\mathbf{M}_{i,j}=0) (\widetilde{x}_{i,j} - x_{i,j})^2,
\end{equation}
where $\mathds{1}(\cdot)$ is the indicator function, $\widetilde{x}_{i,j}$ and $x_{i,j}$ are the reconstructed value and original value of $j$-th sensor in $\widetilde{\mathbf{x}}_i$ and $\mathbf{x}_i$ respectively, and $\mathbf{M}_{i,j}$ is the element in the $i$-th row and the $j$-th column of $\mathbf{M}$.
Thus, we can pre-train the sEMG Intrinsic Pattern Capture via the above strategy to obtain well-initialized model parameters for the downstream task.
In practice, we empirically set the masking proportion $r$ as $0.15$ and the average length of masked segments as $3$.
The illustration of the pre-training procedure is in Figure \ref{overview} (a).

\subsection{Long-term and Short-term Decoding}
Then, based on the pre-trained sEMG Intrinsic Pattern Capture, we develop two decoder heads to further extract the long-term and short-term dependency on the signal sequences, respectively. 
Intuitively, both the long-term and short-term information on signal sequences are significant in the gesture recognition problem. Long-term information refers to the global context of an sEMG sequence, which provides the overall structure of a signal to help the interpretation of the gesture. Short-term information refers to the movement signal in a short time interval of the whole sequence, which can provide specific local characteristics for accurate recognition when the overall structures of sEMGs are ambiguous. For example, distinguishing between Index Finger Extension (IFE) and Middle Extension (ME) movements requires a closer examination of the local signal changes in sEMG, whereas differentiating gestures with large variations, such as hand gestures and wrist gestures, necessitates a focus on the global sEMG information.


\subsubsection{Preserving Long-term sEMG Signal}
Given the hidden output $\mathbf{X}^{(L)} = [\mathbf{x}_1^{(L)}, \mathbf{x}_2^{(L)}, ..., \mathbf{x}_t^{(L)}]$ of a sEMG signal sequence, we first build a long-term decoder to extract the long-term dependency on the complete output. Specifically, the long-term decoder is defined as a multi-head self-attention layer:
\begin{equation}
\label{multihead}
MultiHead\_L\left(\mathbf{X}^{(L)}\right) = Concat \left(h_1, \ldots, h_d\right)\mathbf{W}^O,
\end{equation}

\begin{equation}
\label{head}
\left\{h_i\right\}_{i=0}^{d} = \left\{Attention\left(\mathbf{Q}_i,\mathbf{K}_i,\mathbf{V}_i\right)\right\}_{i=0}^{d},
\end{equation}

\begin{equation}
\label{attention}
Attention\left(\mathbf{Q}_i,\mathbf{K}_i,\mathbf{V}_i\right) = Softmax\left(\frac{\mathbf{Q}_i\mathbf{K}_i^T}{\sqrt{h}}\right)\mathbf{V}_i,
\end{equation}

\begin{equation}
where  \  \mathbf{Q}_i = \mathbf{X}^{(L)} \mathbf{W}_i^Q, \
 \mathbf{K}_i = \mathbf{X}^{(L)} \mathbf{W}_i^K, \
 \mathbf{V}_i = \mathbf{X}^{(L)} \mathbf{W}_i^V,
\end{equation}
where $\{\mathbf{W}_i^Q, \mathbf{W}_i^K, \mathbf{W}_i^V\}_{i=0}^{d}\in \mathbb{R}^{h\times h}$ are parameter matrices and $d$ is the number of attention heads. $Concat(\cdot)$ represents the concatenate operation. $\mathbf{W}^O\in \mathbb{R}^{dh\times h}$ is the output parameter matrix to transform the concatenated outputs of $d$ attention heads.
Then, the long-term sEMG embeddings $\mathbf{H}^{l} \in \mathbb{R}^{t\times h}$ is obtained by 
$\mathbf{H}^{l} = MultiHead\_L(\mathbf{X}^{(L)})$.
Through the self-attention layer, the global context signal information is collected to the embeddings of $t$ timesteps with different attention weights. 

\subsubsection{Preserving Short-term sEMG Signal}
To model the local context information within a short time interval, we introduce a slide-window self-attention layer to extract the short-term dependency on the signal outputs. Similarly, we stack multiple attention heads and calculate the attention of context signals to weighted sum them up into the final representations. The difference is that, for each time step, we only calculate the attention of its nearest $w$ context. Specifically,
we can rewrite the $\mathrm{Attention}(\cdot)$ in Eq.\ref{attention} as:

\begin{equation}
\small    Attention\_S(\mathbf{Q},\mathbf{K},\mathbf{V}) = 
    \left[Softmax\left(\frac{\mathbf{Q}_i {\mathbf{K}_i^w}^T}{\sqrt{h}}\right){\mathbf{V}_i^w}\right]_{i=1}^{t},
\end{equation}

\vspace{-4mm}
\begin{equation}
\small    
\left[\mathbf{K}_i^w\right]_{i=1}^{t} = Unfold(\mathbf{K}, w),\
 \left[\mathbf{V}_i^w\right]_{i=1}^{t} = Unfold(\mathbf{V}, w),
\end{equation}
where $w$ is the sliding windows size, $\mathbf{Q}_i\in \mathbb{R}^h$ as the $i$-th query is the $i$-th row of $\mathbf{Q}$, $\mathbf{K}_i^w \in \mathbb{R}^{w\times h}$ and $\mathbf{V}_i^w \in \mathbb{R}^{w\times h}$ are the keys and values in a window around the $i$-th query. 
For the key and value matrix, we utilize the $Unfold(\cdot)$ operation to generate the slide windows for each timestep. Noted that to avoid confusion, we omit the index of attention head in the above equation. 

Thus, by stacking multiple sets of parameters in $Attention\_S(\cdot)$ to constitute different attention heads, we can obtain the short-term sEMG embeddings $\mathbf{H}^{s} \in \mathbb{R}^{t\times h}$  by 
$\mathbf{H}^{s} = MultiHead\_S(\mathbf{X}^{(L)})$.
Using slide windows, each row in $\mathbf{H}^{s}$ preserves the local context signal information of the corresponding timestep, representing the movement from the past $w/2$ timesteps to the next $w/2$ timesteps.

Unlike LST-EMG-Net \cite{zhang2023lst} and Focal Transformer \cite{yang2021focal}, which process Long Term Feature and Short Term Feature sequentially, we handle them in parallel. This ensures that features processed earlier in a sequential manner are not overlooked.

\subsubsection{Fusion}
Obtained the long-term embeddings $\mathbf{H}^{l}$ and the short-term embeddings $\mathbf{H}^{s}$ of an sEMG signal sequence, we first concatenate them in terms of the hidden dimension, then introduce a 1-D convolution to summarize the $t$-step sEMG embedding sequence into the final sEMG representation, which is fed into a \textit{Feed Forward Layer} with a \textit{Sigmoid Layer} to obtain the final classification probability of which gesture the sEMG belonging to, which can be written as:
$    \widehat{\mathbf{y}} = \sigma(FC(\mathbf{u}^T\cdot [\mathbf{H}^l:\mathbf{H}^s])),$
where $FC(\cdot)$ is a two-layer fully connected
layer, $\sigma(\cdot)$ is the activation function,
and $\widehat{\mathbf{y}}\in \mathbb{R}^C$ is the output classification probability of the sEMG signals.

\subsection{Asymmetric Optimization}

As the common use in multi-label classification, we reduce the gesture recognition problem into a series of binary classification tasks. 
However, we consider two tricky problems that exist in the above model optimization.
(1) As the sampled signals are usually unstable over time, making the samples of the sEMG signal sequence may be critically biased. Some samples with strong signals are easily predicted, while many samples with fuzzy signals are hard to predict. 
(2) As we set 17 classes for the sEMG signals, and each class contains a comparable number of samples. Thus, each class contains, on average, many more negative samples than positive ones. This imbalance may make the model eliminate the gradients from the positive samples in the optimization process, resulting in poor accuracy.
Realizing this, we introduce the Asymmetric loss~\cite{ridnik2021asymmetric} for the gesture classification task. Asymmetric loss is a variant of Focal loss.
(1) It uses focusing parameters to reduce the contribution of easily predicted samples and make the model optimization focus on hard samples; 
(2) It further introduces asymmetric focusing parameters and asymmetric probability shifting to down-weight the contribution from massive easy negatives and emphasizes the contribution of positive samples.
Thus, we define the loss function as follows: 

\begin{equation}
{\small
\begin{aligned}
\mathcal{L}_{\method{}} = -\sum_{i=1}^{N}\sum_{j=1}^{C} &\left({y}_{i,j}\left(1-\widehat{y}_{i,j}\right)^{\gamma^+}\log\left(\widehat{y}_{i,j}\right) +\right. \\
&\left.\left(1-{y}_{i,j}\right)\left(\widehat{y}^m_{i,j}\right)^{\gamma^-}\log\left(1-\widehat{y}^m_{i,j}\right)\right),    
\end{aligned}}
\end{equation}
\begin{equation}
\small
\widehat{y}^m_{i,j} = max\left(\widehat{y}_{i,j}-m,0\right),
\end{equation}
where ${y}_{i,j}$ and $\widehat{y}_{i,j}$ is the ground-truth and probability of the $i$-th sEMG signal sequence belonging to the $j$-th gesture. $\left(1-\widehat{y}_{i,j}\right)^{\gamma^+}$ and $\left(\widehat{y}^m_{i,j}\right)^{\gamma^-}$ are two terms to make the weights of hard predicted samples bigger than those easily predicted samples, $\gamma^+$, and $\gamma^-$ are two focusing parameters and $\gamma^+ > \gamma^-$ lead to asymmetric focusing that help the optimization pay more focus on positive samples of each class.  $\widehat{y}^m_{i,j}$ is the shifted probability and $m$ is shifting margin. The probability shifting for negative samples encourages the optimizer to further reduces their contribution.

\section{Experiment}
\subsection{Settings}
\paragraph{Implementation Details}
STET is implemented in PyTorch \cite{paszke2019pytorch}  and is trained using one RTX 3090 GPU.
During training, we use the RAdam \cite{liu2019variance}, which is a theoretically sound variant of the Adam optimizer with a weight decay of 1e-3. We pre-train on GRABMyo for 20 epochs using a fixed learning rate of 1e-4 for the  backbone. In the decoder, we use two layers of full attention in the long-term decoder and two layers of sliding window attention in the short-term decoder. The short-term decoder's window size is 41 and 21, and the window's move step is set to 1. In both the pre-training and fine-tuning periods, we set the batch size to 16. To avoid overfitting, we set drop out to 0.2.\\

\begin{table*}[htbp]

\label{tab:baseline}
\resizebox{\linewidth}{!}{

\begin{tabular}{l|cc|cc|cc|cc|cc}
\toprule
\multirow{2}{*}{Model} 
& \multicolumn{2}{c|}{Single-finger} & \multicolumn{2}{c|}{Multi-finger} &\multicolumn{2}{c|}{Wrist}   & \multicolumn{2}{c|}{Rest}  & \multicolumn{2}{c}{Overall} \\
& \multicolumn{1}{c}{ACC}   & \multicolumn{1}{c|}{STD}   & \multicolumn{1}{c}{ACC}   & \multicolumn{1}{c|}{STD} & \multicolumn{1}{c}{ACC}     & \multicolumn{1}{c|}{STD}   & \multicolumn{1}{c}{ACC}   & \multicolumn{1}{c|}{STD}   & \multicolumn{1}{c}{ACC}   & \multicolumn{1}{c}{STD} \\
\midrule
Asif \textit{et al.} \cite{asif2020performance} & 83.44\% & 0.015 & 83.58\% & 0.013  & 89.40\% & 0.009 & 90.86\% & 0.012 & 85.34\% & 0.014 \\
TCN \cite{tcn} & 78.78\%  &  0.017 & 79.10\%  &  0.018 & 87.27\% & 0.011 & 88.57\%  &  0.017   &   81.50\%      &      0.016        \\
GRU \cite{chen2021semg}                         & 84.45\%                            &  0.015                        &  84.88\%                           &  0.013                     &       90.06\%                      &      0.009                     &       89.42\%                          &   0.019                        &   86.30\%                      &    0.015                   \\
TEMGNet \cite{rahimian2021temgnet} &   77.70\%                          &     0.019                   &   74.00\%                          &   0.029                      &   84.04\%                          &    0.017                 &   87.46\%                              &   0.014                       &      78.02\%                   &    0.019                    \\
Zerveas \textit{et al.} \cite{zerveas2021transformer} &   78.45\%                          &     0.016                  &   77.20\%                          &   0.020                     &   87.28\%                          &    0.016               &   86.76\%                              &   0.017                       &      80.43\%                   &    0.018                    \\
Informer  \cite{zhou2021informer}                      &   86.88\%                          &   0.016                        &   86.54\%                          &   0.017      &     91.90\%                       &    0.011                      &    83.56\%     &  0.024                        &   87.71\%                      &  0.016                       \\
LST-EMG-Net  \cite{zhang2023lst}                      &   87.21\%                          &   0.011                        &   83.16\%                          &   0.012      &     88.36\%                       &    0.018                      &    82.52\%     &  0.021                        &   85.31\%                      &  0.015                       \\
\midrule
TEMGNET+STEM(ours)                       &  84.57\%                           & 0.017                         & 81.23\%                           &   0.022                     &        88.12\%                   &    0.017                      &  88.74\%                             &   0.013   &                   84.07\%                  &      0.017                \\
Informer+STEM(ours)                       &  87.42\%                           & 0.015                          & 88.39\%                           &   0.018                     &        92.07\%                   &    0.013                      &  90.33\%                             &   0.011   &                   89.14\%                  &      0.015                \\
STET                        &  \textbf{88.27\%}                           & 0.014                          & \textbf{89.93\%}                            &   0.015                     &         \textbf{93.77\%}                    &    0.010                      &  \textbf{95.33\%}                              &   0.012                       & \textbf{90.76\%}                    &      0.012                 \\
\bottomrule
\end{tabular}
}
\caption{The gesture classification performance on the Single-finger, Multi-finger, Wrist, Rest, and Overall categories.}
\vspace{-3mm}
\end{table*}

\vspace{-0.5cm}
\subsubsection{Evaluation Metrics}
Following the prior works\cite{Guo2021,Wang2020,chen2021semg,rahimian2021temgnet}, we choose the below metrics to evaluate the model's performance. \textbf{Pearson Correlation Coefficient} (CC) is a widely used measure of the linear relationship between two variables. It ranges from -1 to 1, where a larger CC value indicates greater similarity between the predicted and estimated joint angles curve, indicating improved estimation.
\textbf{Root Mean Square Error} (RMSE) is a common metric for evaluating the deviation between predicted and observed values.  As the range of fluctuations in the curves of different joint angles can vary significantly, it is difficult to evaluate the performance of models using RMSE alone fairly. Normalization of RMSE addresses this issue, resulting in the Normalized RMSE (NRMSE).
\textbf{Average curvature} ($\kappa$) of all points for each joint is used to measure the smoothness of an estimated curve. A smaller $\kappa$ indicates a smoother curve.

\vspace{-0.1cm}
\subsection{Comparison with Baselines}
We compare the  accuracy (ACC) and Standard deviation (STD) between our proposed STET and previous popular sEMG-based gesture recognition methods. 
Specifically, we train the model on the GRABMyo dataset \cite{pradhan2022multi} (the detail of data processing shown in section \ref{dataset} ) and separately report the classification results on the categories of Single-finger gestures, Multi-finger gestures, Wrist gestures, Rest, and the overall results.
The ratio of the training set to the test set for each gesture is 5 to 2.


From the experimental results, we can observe that STET consistently performs best on four categories of gestures and overall data.
In particular, STET and \cite{zerveas2021transformer} both use Transformer-based encoders.
While in the decoder part, STET introduces both the short-term and long-term decoder rather than the fully connected layers used in \cite{zerveas2021transformer}.
As a result, the overall accuracy of STET is improved from 80.43\% to 90.76\% compared to \cite{zerveas2021transformer}. This is because the proposed long-term and short-term decoupling module can extract both the global and fine-grained dependency on the signal dependency and thus can learn better sEMG representations.


Among the transformer-based methods, Informer and STET performed best, with accuracy rates of 87.71\% and 90.76\%, respectively. Informer relies heavily on max pooling layers to aggregate features, leading to the relative weakness in extracting some short-term features. STET enhances accuracy and stability by strengthening the short-term feature extraction. The improvement is remarkable on the Rest gestures, where the accuracy improves from 83.56\% to 95.33\%.  Furthermore, after incorporating our designed short-term encoder into Informer, its accuracy rate increased from 87.71\% to 89.14\%, and the classification accuracy for Rest gestures improved from 83.56\% to 90.33\%. 

Note that the Rest category of gesture, as the resting state of devices such as interactive bracelets, is the most frequent gesture that appears in the signals. 
Thus, the stability of its prediction plays a significant role in the problem.
STET achieves the highest and most stable results on the prediction of the Rest category compared with all baselines, indicating the robustness of STET.

\subsection{Ablation Studies}
To validate the effects of the unsupervised sEMG Intrinsic Pattern Capture (EIPC), Long-term decoder, Short-term decoder, Fuse strategy, and loss function. 
We designed variants of STET and reported their results in Table \ref{tab:my-table}.

First, we can observe that, with the unsupervised EIPC, the accuracy of the transformer and STET is improved by 0.60\% and 1.12\% compared with training from scratch. This suggests that unsupervised EIPC can aid in discerning additional data features, such as the inherent variability in sEMG, without the requirement for new samples or extra annotations. Significantly, this process circumvents the need for external data, thus preserving user privacy in the context of data acquisition and processing.
Replacing the fully connected layer of the transformer's decoder with the long-term decoder or short-term decoder,
the performance is improved by 1.16\% and 1.39\%, respectively. 
Furthermore, the performance is comparable when using the long-term decoder or short-term decoder alone, indicating that the two kinds of features may play different significant roles in the sEMG signal recognition, and the short-term cannot be ignored.
Most importantly, when we used our designed fuse module to combine long-term and short-term features, the accuracy further improved by 2.46\%. This suggests that the two decoders are complementary and that enhancing short-term features is necessary on the basis of the long-term decoder. We employed Asymmetric Loss (ASL) to drive the model's focus on difficult samples. Compared to simply using the Cross-Entropy Loss (CEL), the accuracy improved by 0.73\%, indicating the effectiveness of ASL.

\begin{table}[tbp]

\resizebox{\linewidth}{!}{
\begin{tabular}{c|c|c|c|c|c|c|c}
\toprule
Transformer & EIPC & LT & ST & Fusing & CEL &ASL & ACC \\
\midrule 
 \checkmark &   &  &  &  & \checkmark & & 85.73\%  \\
 \checkmark & \checkmark  &  &  &  & \checkmark& & 86.33\% \\
 \checkmark & \checkmark & \checkmark  &  &  & \checkmark & & 88.02\% \\
 \checkmark & \checkmark &             & \checkmark &  & \checkmark & & 87.72\% \\
 \midrule
 \checkmark & \checkmark & \checkmark  & \checkmark & \checkmark & \checkmark & & 89.37\% \\
  \checkmark &  & \checkmark  & \checkmark & \checkmark &  &\checkmark & 89.42\% \\
 \checkmark & \checkmark & \checkmark  & \checkmark & \checkmark & & \checkmark & 90.54\% \\
\bottomrule 
\end{tabular}
}
\caption{The results of ablation study. EIPC: sEMG Intrinsic Pattern Capture; LT: Long-Term decoder; ST: Short-Term decoder; CEL: Cross Entropy Loss; ASL: ASymmetric Loss.}
\label{tab:my-table}
\vspace{-5mm}
\end{table}

\begin{figure}[tb] 
\centering 
\includegraphics[width=0.47\textwidth]{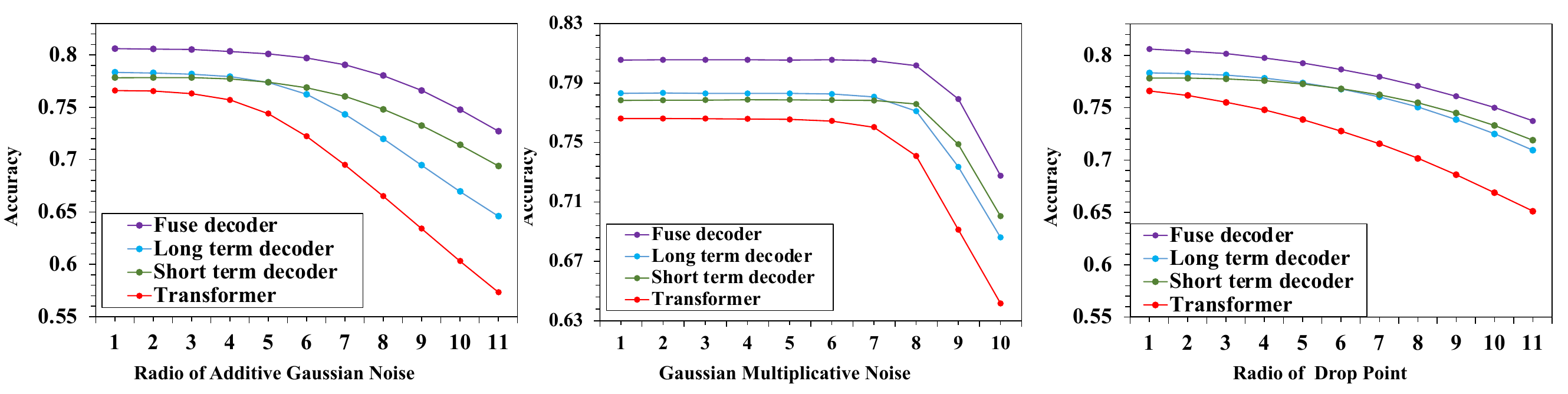} 
\vspace{-2mm}
\caption{Accuracy versus Noise Intensity Curve.} 
\label{nosiy} 
\vspace{-3mm}
\end{figure}

\begin{table}[htb]
\centering

\resizebox{\columnwidth}{!}{%
\begin{tabular}{|c|c|c|c|c|}
\hline
Backbone & In STET framework & AG noise & MG noise & Signal loss \\
\hline
Transformer & No & 25\% & 16\% & 14\% \\
Transformer & Yes & 10\% & 10\% & 8\% \\
\hline
Informer & No & 11\% & 9\% & 26\% \\
Informer & Yes & 9\% & 8\% & 17\% \\
\hline
\end{tabular}
}
\caption{Drop rates of accuracy calculated by \(drop\_rate=\frac{ACC_{raw}-ACC_{noise}}{ACC_{raw}}\). AG: Additive Gaussian noise, MG: Multiplicative Gaussian noise}
\label{tab:noise_droprate}
\vspace{-0.4cm}
\end{table}

\vspace{-2mm}
\subsection{Robustness Analysis}
To verify the robustness of the model, we only used high-quality data collected in the lab to train the model and added different types of noise (Additive Gaussian noise, Multiplicative Gaussian noise, and signal loss) during validation to simulate complex scenarios that might be encountered in real situations.

Additive noise typically refers to thermal noise, which is added to the original signal. This type of noise exists regardless of the presence of the original signal and is often considered the background noise of the system in sEMG acquisition. Multiplicative noise is generally caused by channel instability, and it has a multiplicative relationship with the original signal. Also, we simulated signal loss during transmission by randomly setting a portion of the signals to zero.

Figure \ref{nosiy} illustrates the influence exerted by three distinctive noise categories, namely additive noise, multiplicative noise, and signal loss, on the accuracy of the proposed model. The model using only the short-term decoder is less affected by noise compared to the long-term version. This relative robustness of the short-term decoder is potentially attributable to its unique capability to mitigate the global impact of noise by virtue of a sliding window multiple-sampling scheme, which effectively confines the sphere of noise impact. The model that integrates both long-term and short-term characteristics persistently outperforms models that rely on only one. This highlights the significant effectiveness of the integrating process in dealing with noise-induced interference. As depicted in Table~\ref{tab:noise_droprate}, it is evident that both Transformer and Informer models demonstrate a notable enhancement in noise resistance when their decoders are replaced with the design from STET.

\begin{figure}[tbp] 
\centering 
\includegraphics[width=0.47\textwidth]{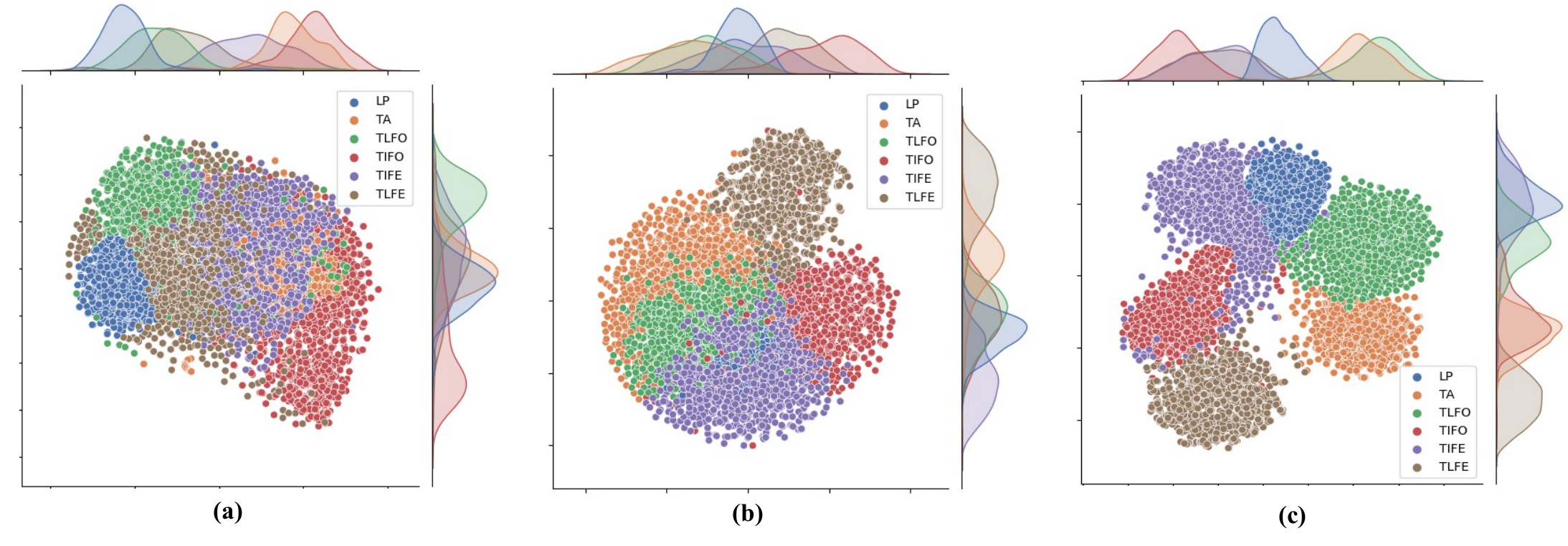} 
\vspace{-3mm}
\caption{Visualization of (a) the long-term sEMG embeddings, (b) the short-term sEMG embeddings, and (c) the fused sEMG embeddings for gesture recognition. Note that we color each sample by its classes.}  
\vspace{-0.6cm}
\label{clustering} 
\end{figure}

\subsection{Visualizations}
To demonstrate the distinction, we first obtained STET's long-term, short-term, and fuse embeddings. The embeddings with dimensions $(N,T,H)$ were then flattened to $(N,T*H)$ and separately projected in 2D by t-SNE, the result shown in Figure \ref{clustering}. We colored each node by category for the illustration. As shown in Figure \ref{clustering}, The classification boundary generated by the long-term feature and the short-term feature is a significant difference, indicating that  the long-term and short-term features are capable of recognizing different types of gestures. This further suggests that the two features are complementary in data representation. For example, short-term embedding can distinguish TA gesture and TIFO gesture very well, but TA gesture and TIFO gesture will be confused in long-term embedding. Meanwhile, long-term embedding can distinguish LP gesture and TIFE gesture very well, but short-term embedding will confuse them. As shown in Figure \ref{clustering}(c), after the fusion of the two types of features, the classification interface is wider, and the confusion points are significantly reduced, which indicates that the fusion module can effectively complement the strengths of the two types of features.

\begin{table}[t]
\centering
\resizebox{\linewidth}{!}{
\begin{tabular}{ccccc}
\toprule
Model & PCC    & NRMSE  & $\kappa$ & Time Cost/epoch(s)     \\
\midrule
LSTM  & 0.779  & 0.096  & 0.581 & 26.36 \\
TCN   & 0.833 & 0.088 & 1.533 & \textbf{3.62} \\
BERT  & 0.867 & 0.077 & 1.571 & 4.95 \\
sBERT-OHME & 0.869 & 0.076 & 0.532 & 4.96 \\
STET  & \textbf{0.877} & \textbf{0.073} & \textbf{0.522} & 6.83\\
\bottomrule
\end{tabular}}

\caption{Comparison of STET with Other models in Predicting Joint Angle for Fingers.}
\label{tab:regress-table1}
\vspace{-6mm}
\end{table}

\subsection{Regression: Hand Joint Angles Prediction}
STET can conveniently handle regression tasks by changing the loss function to mean squared error (MSE) loss. Continuous motion estimation extracts continuous motion information, such as joint angles and torques, from sEMG signals. Since continuous motion estimation requires outputting subtle variations of the movement at each time instant, the local signal variations are particularly important for this type of estimation. In this section, we have re-selected the most competitive models known for sEMG-based joint angle prediction as the baseline and tested the performance of STET on the regression task of predicting the main 10 joint angles for fingers using the Ninapro DB2 \cite{atzori2014electromyography}  dataset. As shown in Table \ref{tab:regress-table1}, STET achieved the best performance in PCC, NRMSE, and $\kappa$, indicating that the joint angle curve predicted by STET is more in line with the real curve and has less abnormal fluctuations, which will significantly improve the user's interactive experience. 
In terms of training time, due to the addition of the short-term decoder, its training speed is slightly slower than BERT but still within an acceptable range.

\section{Conclusion}
Current sEMG-based gesture recognition models usually fail to handle various noisy and distinguish similar gestures, especially in non-laboratory settings. In this paper, we found using  short-term information and  self-supervised EIPC mitigates this issue.  Therefore, we proposed STEM for capturing local signal changes and enhancing noise resistance. The STEM is easily deployable and serves as a plug-in that can potentially be applied to most time series deep learning models.  According to our experimental results, our method significantly improved performance for both classification and regression tasks in sEMG, and the model's ability to resist signal loss, Gaussian additive noise, and Gaussian multiplicative noise was clearly improved. This will further drive the practical application of sEMG in VR, AR, and other human-computer interaction scenarios.

\bibliographystyle{named}
\bibliography{ijcai24}

\end{document}